\RequirePackage[mathlines]{lineno}
\documentclass[prd,twocolumn,amsmath,amssymb]{revtex4}
\usepackage{overpic,graphicx}
\usepackage{dcolumn}
\usepackage{bm}
\usepackage{rotating}
\usepackage{subfigure}
\usepackage{color}
\usepackage{multirow}
\usepackage{epstopdf}
\usepackage{mathrsfs}
\usepackage{comment}
\usepackage{booktabs}
\usepackage{defs_LamcPEtap}
\usepackage{gensymb}
\usepackage{slashed}
\usepackage{hyperref}

\renewcommand\tablename{\rm Table}

\begin{document}

\title{\bf \boldmath
Search for a massless dark photon in $\Lambda^{+}_{c}\to p \gamma^{\prime}$ decay
}

\author{
\begin{small}
\begin{center}
M.~Ablikim$^{1}$, M.~N.~Achasov$^{11,b}$, P.~Adlarson$^{70}$, M.~Albrecht$^{4}$, R.~Aliberti$^{31}$, A.~Amoroso$^{69A,69C}$, M.~R.~An$^{35}$, Q.~An$^{66,53}$, X.~H.~Bai$^{61}$, Y.~Bai$^{52}$
% ...
\\
\vspace{0.2cm}
(BESIII Collaboration)\\
\vspace{0.2cm} {\it
$^{1}$ Institute of High Energy Physics, Beijing 100049, People's Republic of China\\
% ...
$^{4}$ Bochum Ruhr-University, D-44780 Bochum, Germany\\
% ...
$^{11}$ G.I. Budker Institute of Nuclear Physics SB RAS (BINP), Novosibirsk 630090, Russia\\
% ...
$^{31}$ Johannes Gutenberg University of Mainz, Johann-Joachim-Becher-Weg 45, D-55099 Mainz, Germany\\
% ...
$^{35}$ Liaoning Normal University, Dalian 116029, People's Republic of China\\
% ...
$^{52}$ Southeast University, Nanjing 211100, People's Republic of China\\
$^{53}$ State Key Laboratory of Particle Detection and Electronics, Beijing 100049, Hefei 230026, People's Republic of China\\
% ...
$^{61}$ University of Jinan, Jinan 250022, People's Republic of China\\
% ...
$^{66}$ University of Science and Technology of China, Hefei 230026, People's Republic of China\\
% ...
$^{69}$ University of Turin and INFN, (A)University of Turin, I-10125, Turin, Italy; (B)University of Eastern Piedmont, I-15121, Alessandria, Italy; (C)INFN, I-10125, Turin, Italy\\
$^{70}$ Uppsala University, Box 516, SE-75120 Uppsala, Sweden\\
% ...
$^{b}$ Also at the Novosibirsk State University, Novosibirsk, 630090, Russia\\
% ...
}
\end{center}
\vspace{0.4cm}
\end{small}
}

\begin{abstract}
A search for a massless dark photon $\gamma^{\prime}$ is conducted 
using 4.5 $\invfb$ of $\ee$ collision data collected at center-of-mass energies
between 4.600 and 4.699 $\gev$ with the BESIII detector at BEPCII.
No significant signal is observed, and the upper limit on
the branching fraction $\mathcal{B}(\LamCPI)$ is determined to be $8.0\times 10^{-5}$ at 90\% confidence level.

\end{abstract}


\maketitle

\oddsidemargin  -0.2cm
\evensidemargin -0.2cm

\section{Introduction}

The flavor changing neutral current (FCNC) transitions of
the charmed baryon $\LamC$ are of great interest since they can
provide indications for physics beyond the Standard Model
(SM). In the framework of the SM, FCNC transitions are
strongly suppressed by the Glashow, Iliopoulos and Maiani
(GIM) mechanism~\cite{Glashow:fcnc} in the charm sector.
The SM predictions for the branching fractions (BFs) of FCNC decays in the charm sector
are less than $10^{-9}$~\cite{Tanabashi:charmFCNC}.
The minimal super-symmetric SM with R-parity violation~\cite{Aulakh:model1}
and the two-Higgs-doublet model~\cite{Glashow:model2} predict the BFs of the
same FCNC decays to be two to three orders of magnitude larger.
Observation of a FCNC decay with the current experimental sensitivity would
imply new physics beyond the SM.

Models of new physics beyond the SM may have a dark sector containing an extra Abelian
gauge group, $U(1)_D$, under which all the SM fields are singlets.
This symmetry may be broken spontaneously or may remain unbroken,
causing the associated gauge boson, the dark photon, to acquire 
a mass or remain massless. 
These possibilities have received a great deal of attention
in recent decades~\cite{dark:mode1,Dobrescu:dark1,Dobrescu:dark2,dark:mode2,Ying:darkphoton,Ackerman:intre1,Pan:intre2,Barger:intre3,Chiang:intre4,Batley:intre5,Pospelov:intre6}.
If $U(1)_D$ remains unbroken, there is
always a linear combination of the dark and SM Abelian
gauge fields which does not have renormalizable couplings
to SM members and which can be identified with the
massless dark photon ($\gamma^{\prime}$)~\cite{dark:mode1,Dobrescu:dark1}.
While it has no direct interactions with SM fermions, 
the $\gamma^{\prime}$ can still exert influence on the SM via
higher-dimensional operators generated by loop diagrams
involving particles that are charged under $U(1)_D$ and also
coupled to SM fields~\cite{Dobrescu:dark1, Dobrescu:dark2, dark:mode2}.

\begin{figure}[!htp]
    \begin{center}
        \includegraphics[width=0.3\textwidth]{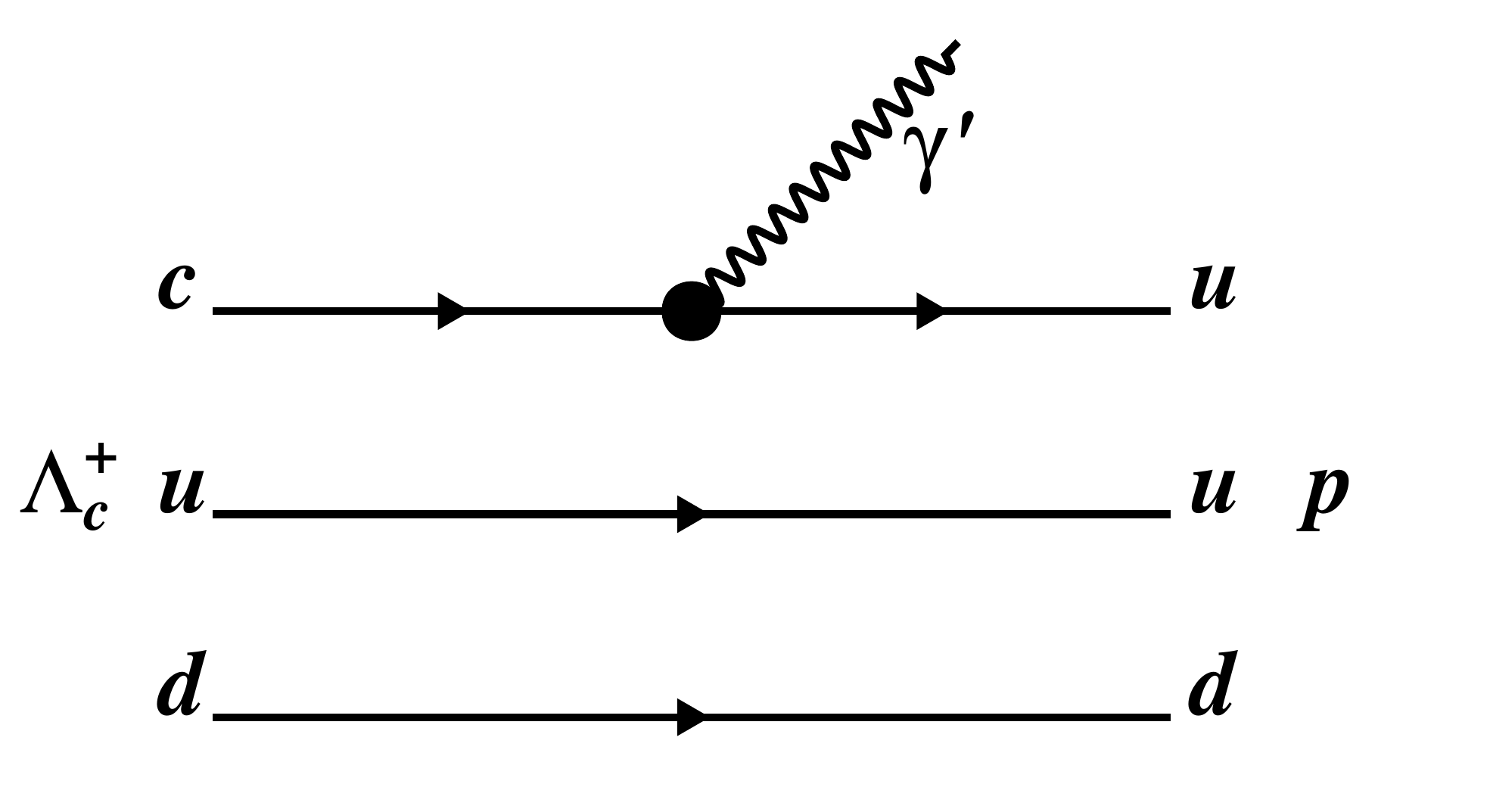}
    \end{center}
    \caption{ Feynman diagram for $\LamCPI$.
      } \label{fig:feynman_pinvisible}
\end{figure}

In experiment, LHCb reported the evidence for the breaking of
lepton universality in bottom-quark FCNC decays to
charged lepton pairs with a significance of $3.1\sigma$~\cite{LHCb:lepvio}. 
As a complementary study, we concentrate on FCNC effects arising from 
the dark photon with the $c$ and $u$ quarks, where the missing energy 
due to the dark photon is the feature of the signal processes. 
BESIII has searched for the invisible signals within various hadron decays, 
including $\eta/\eta^\prime \to \rm{invisible}$~\cite{BESIII:etainv}, $\omega/\phi\to \rm{invisible}$~\cite{BESIII:omegainv}, 
$\Lambda\to \rm{invisible}$~\cite{BESIII:Laminv} and $J/\psi\to \gamma + \rm{invisible}$~\cite{BESIII:Jpsiinv},
and no significant signals are observed.
However, this has never been probed in the charmed baryon sector. 
The two-body charmed baryon decay potentially offers a competitive window to access
$c \to u \gamma^{\prime}$, which gives rise to 
the FCNC decays of charmed baryon into a lighter baryon plus
missing energy carried away by the massless dark photon.
Figure~\ref{fig:feynman_pinvisible} presents a typical Feynman diagram of $\LamCPI$.
It is found that the BFs of some charmed baryon decays are allowed to be
as high as a few times $10^{-5}$~\cite{Tandean:cFCNC} . Such
BFs are likely to be within the sensitivity reaches of some ongoing experiments like BESIII.

This paper presents an experimental search for a massless dark photon in $\LamCPI$ decay using 4.5~$\invfb$ of 
$\ee$ collision data collected with the BESIII detector at seven center-of-mass (CM) energies between 4.600 and 4.699 GeV. 
The CM energies and the integrated luminosities for each energy point are listed in \tablename~\ref{tab:data_sets}~\cite{BESIII:lumi0, BESIII:lumi1, BESIII:lumi2}.
Taking advantage of the $\LCpair$ production just above the mass threshold 4572.92~MeV,
a double-tag (DT) approach~\cite{MarkIII:DT} is implemented.
Throughout the text, the charge conjugate states are always implied.

\begin{table}[!htbp]
  \begin{center}
    \caption{The CM energies and the integrated luminosities ($\mathcal{L}_{int}$) 
             for each energy point.
             The first and the second uncertainties are statistical and systematic, respectively.}
    \begin{tabular}{ c  c}
      \hline
      \hline
         $\sqrt s$ (MeV) &  $\mathcal L_{int}$ (\ipb)  \\
      \hline
              4599.53 $\pm$ 0.07 $\pm$ 0.74   &~~  586.90  $\pm$ 0.10 $\pm$ 3.90 \\
              4611.86 $\pm$ 0.12 $\pm$ 0.32   &~~  103.83  $\pm$ 0.05 $\pm$ 0.55 \\
              4628.00 $\pm$ 0.06 $\pm$ 0.32   &~~  521.52  $\pm$ 0.11 $\pm$ 2.76 \\
              4640.91 $\pm$ 0.06 $\pm$ 0.38   &~~  552.41  $\pm$ 0.12 $\pm$ 2.93 \\
              4661.24 $\pm$ 0.06 $\pm$ 0.29   &~~  529.63  $\pm$ 0.12 $\pm$ 2.81 \\
              4681.92 $\pm$ 0.08 $\pm$ 0.29   &~   1669.31 $\pm$ 0.21 $\pm$ 8.85 \\
              4698.82 $\pm$ 0.10 $\pm$ 0.39   &~~  536.45  $\pm$ 0.12 $\pm$ 2.84 \\
      \hline\hline
    \end{tabular}
      \label{tab:data_sets}
  \end{center}
\end{table}

\section{BESIII Detector and Monte Carlo simulation}

The BESIII detector~\cite{Ablikim:2009aa} records symmetric $e^+e^-$ collisions
provided by the BEPCII storage ring, which operates with a
peak luminosity of $1\times10^{33}$~cm$^{-2}$s$^{-1}$ in the CM energy range 
from 2.0 GeV to 4.9~GeV. BESIII has collected large data samples at these energy regions~\cite{Ablikim:2019hff}.
The cylindrical core of the BESIII detector covers 93\% of the full solid angle and
consists of a helium-based multilayer drift chamber~(MDC), a plastic scintillator time-of-flight
system~(TOF), and a CsI(Tl) electromagnetic calorimeter~(EMC),
which are all enclosed in a superconducting solenoidal magnet providing a 
{\spaceskip=0.2em\relax 1.0 T} magnetic field. The solenoid is supported by an
octagonal flux-return yoke with resistive plate counter based muon
identification modules interleaved with steel.
The charged-particle momentum resolution at $1~{\rm GeV}/c$ is $0.5\%$, 
and resolution of the ionization energy loss in the MDC ($\mathrm{d}E/\mathrm{d}x$)
is $6\%$ for electrons
from Bhabha scattering. The EMC measures photon energies with a
resolution of $2.5\%$ ($5\%$) at $1$~GeV in the barrel (end cap) region. 
The time resolution in the TOF barrel region is
68 ps, while that in the end cap region is 110 ps.
The end cap TOF system was upgraded in 2015 using multi-gap
resistive plate chamber technology, providing a time
resolution of 60 ps~\cite{LiGuo:MRPC}.  
About 13\% of the data (the 4.600 GeV sample) used in the current analysis 
predates this upgrade.  

Monte Carlo (MC) simulated data samples are produced with a {\sc geant4}-based~\cite{Agostinelli:2002hh} MC package, which
includes the geometric description and response of the BESIII detector.
The signal MC samples of $\ee\to\LCpair$ with $\LamCB$ decaying into ten specific tag modes (as described below and listed in \tablename~\ref{tab:yield-st-460})
and $\LamCPI$, which are used to determine the detection efficiencies, are generated for each CM energy using the generator {\sc kkmc}~\cite{Jadach:2000ir} 
incorporating initial-state radiation (ISR) effects and the beam energy spread.
The inclusive MC samples, which consist of $\LCpair$ events, 
charmed meson $D_{(s)}^{(\ast)}$ pair production,
ISR return to the charmonium(-like) $\psi$ states at lower masses,
and continuum processes $e^{+}e^{-}\rightarrow q\bar{q}$ ($q=u,d,s$), are generated to estimate
the potential background.
Decay modes as specified in the Particle Data Group (PDG)~\cite{PDG:2020} are modeled with 
{\sc evtgen}~\cite{Lange:2001uf, Ping:2008zz},
and the remaining unknown decays are modeled with {\sc lundcharm}~\cite{Chen:2000tv,YANG:2014}.
Final state radiation~(FSR) from charged final state particles is incorporated using
{\sc photos}~\cite{Richter-Was:1992hxq}.

\section{METHODOLOGY}

A DT approach~\cite{MarkIII:DT} is implemented to search for $\LamCPI$. A data sample of $\LamCB$ baryon,
referred to as the single-tag (ST) sample, is reconstructed with ten exclusive hadronic decay modes,
as listed in \tablename~\ref{tab:yield-st-460}. 
The subset of those events in which a signal decay $\LamCPI$ candidate 
is reconstructed in the system recoiling against
the $\LamCB$ candidate are denoted as DT candidates. 
The $\LamC$ decay BF is determined as

\begin{equation} \label{eq:br}
  \mathcal{B}(\LamCPI)=\frac{N_{\mathrm{obs}} - N_{\mathrm{bkg}}} {\sum_{ij} N_{ij}^{\mathrm{ST}}\cdot (\epsilon_{ij}^{\mathrm{DT}}/\epsilon_{ij}^{\mathrm{ST}}) },
\end{equation}
where the $N_{\mathrm{obs}}$ is the number of observed events in the signal region from data, and
$N_{\mathrm{bkg}}$ is the number of estimated background as explained explicitly in Section~\ref{sec:bkg}.
The subscripts $i$ and $j$ represent the ST modes and the data samples at different
CM energies, respectively. The parameters $N_{ij}^{\mathrm{ST}}$, $\epsilon_{ij}^{\mathrm{ST}}$
and $\epsilon_{ij}^{\mathrm{DT}}$ are the ST yields, ST and DT detection efficiencies, respectively.

\section{ST Event Selections} \label{sec:single-tag}

Charged tracks detected in the MDC are required to be within a polar angle ($\theta$)
range of $|\!\cos\theta| < 0.93$, where $\theta$ is defined with respect to the positron beam direction.
For prompt tracks not from $\Ks$ and $\bar{\Lambda}$ decays, 
the distances of the closest approach to the interaction point (IP) are required to
be within $\pm$10~cm along the beam direction and 1~cm in the plane perpendicular to the beam (referred to as tight track hereafter).
The particle identification (PID) is implemented
by combining measurements of the $\mathrm{d}E/\mathrm{d}x$ and the flight time in the TOF.
Every charged track is assigned a particle type of pion, kaon or proton,
by choosing the type with the highest probability. 

Photon candidates are selected from showers reconstructed in the EMC. 
The deposited energy of each shower must be
more than 25~MeV in the barrel region ($|\!\cos\theta| \le 0.80$) or more than 50~MeV in the end cap region
($0.86 \le |\!\cos\theta| \le 0.92$). To suppress electronic noise and showers unrelated to the event,
the difference between the EMC time and the event start time is required to be within (0, 700)~ns. The $\pi^0$ candidates are reconstructed from photon pairs with an invariant mass in the range 
(0.115, 0.150)~GeV/$c^2$. To improve the resolution, a kinematic fit is performed constraining the invariant
mass of the photon pair to the known $\pi^0$ mass~\cite{PDG:2020}.  The corresponding $\chi^2$ of the fit must be less than 200. 
The momenta updated by the kinematic fit are used in further analysis.

Candidates for $\Ks$ and $\bar{\Lambda}$ are reconstructed in their decays to $\pi^+\pi^-$ and $\bar{p}\pi^+$, respectively.  Each charged track must have
a distance of closest approach to the IP within $\pm$20~cm along the beam direction (referred to as loose track hereafter).
To improve the signal purity, PID is applied to the proton candidates,
but not the pion candidates.  
A secondary vertex fit is performed to each $\Ks$ or $\bar{\Lambda}$ candidate,
and the momenta updated by the fit are used in the further analysis. 
To keep a high signal efficiency, a $\Ks$ or
$\bar{\Lambda}$ candidate is accepted if the $\chi^2$ of this fit is less than 100.
Furthermore, the decay vertex is required to be separated from the IP by a distance of at least twice the fitted vertex resolution, and the invariant mass must be
within (0.487, 0.511)~GeV/$c^2$ for $\pi^+\pi^-$  or (1.111, 1.121)~GeV/$c^2$ for $\bar{p}\pi^+$.
The $\bar{\Sigma}^0$ and $\bar{\Sigma}^-$ candidates are reconstructed with the $\gamma\bar{\Lambda}$
and $\bar{p}\pi^0$ final states, requiring the invariant masses to lie within 
(1.179, 1.203)~GeV/$c^2$ and (1.176, 1.200)~GeV/$c^2$, respectively.

The ST $\LamCB$ candidates are identified using
the beam constrained invariant mass $M_\mathrm{BC} = \sqrt{\Ebeam^2/c^4 - | \pALC |^2/c^2}$
and energy difference $\dE = E_{\ALamC} - \Ebeam$, where $\Ebeam$ is the beam energy,
$E_{\ALamC}$ and $\pALC$ are the energy and momentum of the $\LamCB$ candidate in the $\ee$ CM frame, respectively.
The $\LamCB$ candidates are required to satisfy the tag-mode dependent $\dE$ requirements,
the asymmetric intervals of which take into account the effects of ISR and
correspond to three times the resolution around the peak, as summarized
in  \tablename~\ref{tab:yield-st-460}. 
If there are more than one candidate satisfying the above requirements for
a specific tag mode, the one with the minimum $|\dE|$ is kept.

\begin{table}[!htbp]
  \begin{center}
  \caption{$\dE$ requirement, the ST yield, and the ST detection efficiency 
           of each tag mode for data sample at  $\sqrt{s}=4.600$~GeV. The uncertainty in the ST yield is statistical only. }
    \begin{tabular}{ p{2cm} c  p{1cm}<{\raggedleft} @{ $\pm$ } p{1cm}<{\raggedright} c }

      \hline
      \hline
            & $\dE$ (MeV) & \multicolumn{2}{c}{$N_{i}^{\mathrm{ST}}$}  & $\epsilon_{i}^{\mathrm{ST}}$(\%)    \\
      \hline
            $\Bpkpi$                       & $(-34,~20)$    &  $6705$     & 90  &  51.0   \\
            $\Bpks$                        & $(-20,~20)$    &  $1268$     & 37  &  56.2   \\
            $\bar{\Lambda}\pi^-$           & $(-20,~20)$    &  $741$      & 28  &  47.7   \\
            $\Bpkpi\pi^0$                  & $(-30,~20)$    &  $1539$     & 57  &  15.4   \\
            $\Bpks\pi^0$                   & $(-30,~20)$    &  $485$      & 29  &  18.4   \\
            $\bar{\Lambda}\pi^-\pi^0$      & $(-30,~20)$    &  $1382$     & 49  &  16.6   \\
            $\Bpks\pi^+\pi^-$              & $(-20,~20)$    &  $512$      & 29  &  19.9   \\
            $\bar{\Lambda}\pi^-\pi^+\pi^-$ & $(-20,~20)$    &  $646$      & 31  &  13.7   \\
            $\bar{\Sigma}^0\pi^-$          & $(-20,~20)$    &  $404$      & 22  &  22.5   \\
            $\bar{\Sigma}^-\pi^+\pi^-$     & $(-30,~20)$    &  $872$      & 38  &  18.1   \\
      \hline\hline
    \end{tabular}
    \label{tab:yield-st-460}
  \end{center}
\end{table}

For the $\LamCB\to\Bpks\pi^0$ ST mode, candidate events with
$M_{\bar p\pi^+} \in (1.100,1.125)$~GeV/$c^2$ and $M_{\bar p\pi^0}\in (1.170,1.200)$~GeV/$c^2$ are vetoed
to avoid double counting with the $\LamCB\to\bar{\Lambda}\pi^-\pi^0$ or
$\LamCB\to\bar{\Sigma}^-\pi^+\pi^-$ ST modes, respectively.
For the $\LamCB\to\bar{\Sigma}^-\pi^+\pi^-$ ST mode,
candidate events with $M_{\pi^+\pi^-}\in (0.490,0.510)$~GeV/$c^2$ and
$M_{\bar p\pi^+}\in (1.110,1.120)$~GeV/$c^2$  are rejected
to avoid double counting with the
$\LamCB\to\Bpks\pi^0$ or $\LamCB\to\bar{\Lambda}\pi^-\pi^0$ ST modes, respectively.
In the $\LamCB\to\Bpks\pi^+\pi^-$ and $\bar{\Lambda}\pi^-\pi^+\pi^-$ selections,
candidate events with $M_{\bar p\pi^+}\in (1.100,1.125)$~GeV/$c^2$ and
$M_{\pi^+\pi^-}\in (0.490,0.510)$~GeV/$c^2$ are rejected, respectively.

\begin{figure}[!htp]
    \begin{center}
        \includegraphics[width=0.42\textwidth]{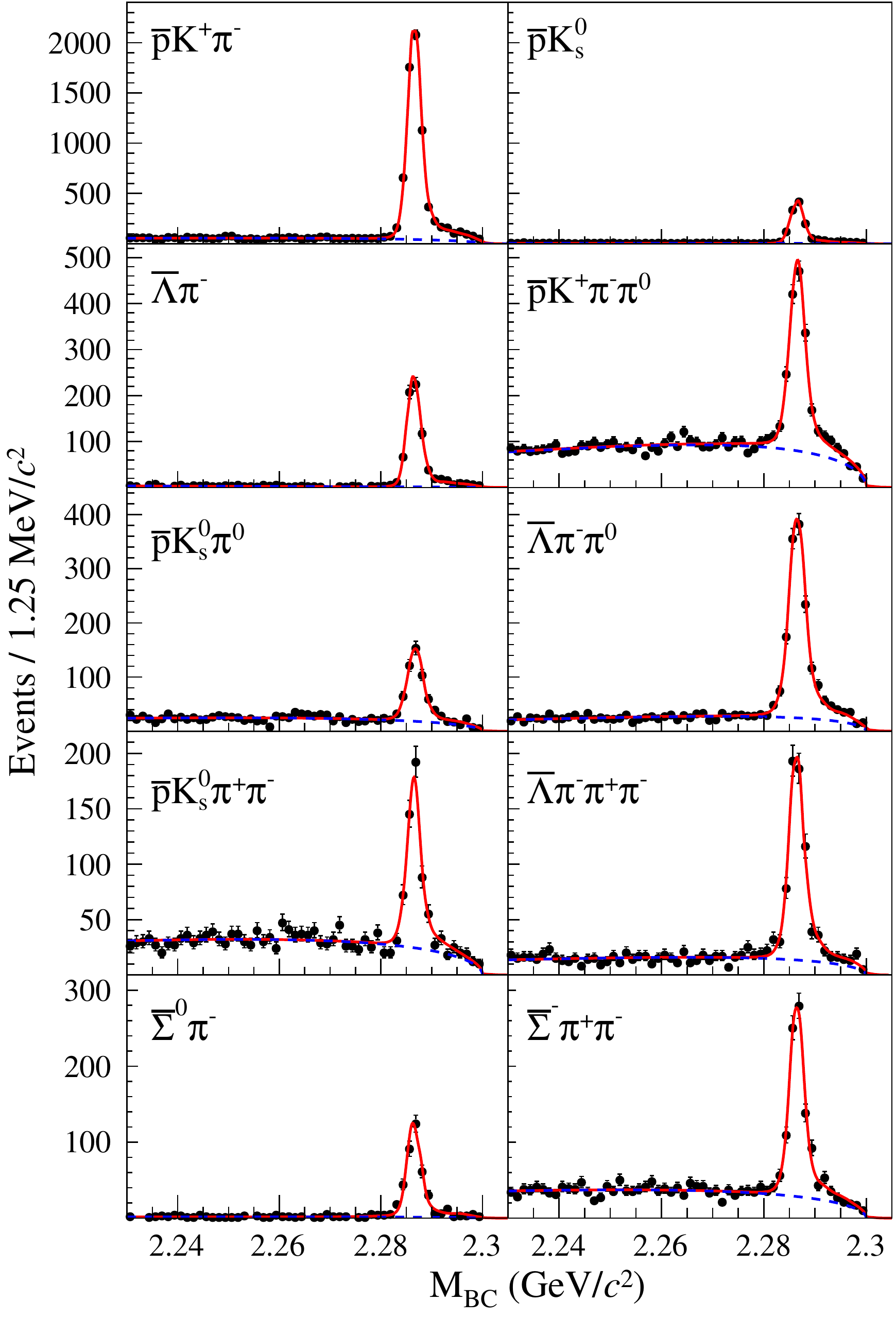}
    \end{center}
    \caption{
      The $\mBC$ distributions of the ST modes for data sample at
      $\sqrt{s}=4.600~\mathrm{GeV}$.
      The points with error bars represent data. The red solid curves indicate the fit results and the blue dashed curves describe
      the background shapes.}
     \label{fig:single-tag-468}
\end{figure}

The $\mBC$ distributions of candidates for the ten ST modes with the data sample at
$\sqrt{s}=4.600~\mathrm{GeV}$ are illustrated in Fig.~\ref{fig:single-tag-468},
where clear $\LamCB$ signals are observed in each mode. 
No peaking background is found using the inclusive MC samples.
To obtain the ST yields,
unbinned maximum likelihood fits on these $\mBC$ distributions are performed,
where the signal shape is modeled with the MC-simulated shape
convolved with a Gaussian function representing for the resolution difference between data and MC simulation,
and the background shape is described by the ARGUS function~\cite{ARGUS:1990hfq}.
The candidates with $\mBC \in (2.275,2.310)$~GeV/$c^2$ are retained for further analysis,
and the signal yields for the individual ST modes are summarized
in \tablename~\ref{tab:yield-st-460}. 
The same procedure is performed for the other six data samples at different CM energies, 
the results can be found in Ref.~\cite{Ablikim:2022SCS} and its supplemental material. 
The sum of ST yields for all data samples at different CM energies is $105244 \pm 384$, where the uncertainty is statistical.

\section{Reconstruction of $p \gamma^{\prime}$ candidates}

The decay $\LamCPI$ is searched for among the remaining tracks 
and showers recoiling against the $\LamCB$ candidates.
Only one tight track is allowed, and it must satisfy the PID criteria 
$\mathcal L$($p$)  $> \mathcal L$($K$) and $\mathcal L$($p$)  $> \mathcal L$($\pi$).
To suppress contamination from long-lifetime particles in the
final state, the candidate events are further required to
be without any unused loose tracks. 
The $\gamma^{\prime}$ does not interact directly with SM fermions and thus it deposits no energy in the EMC.  
Backgrounds containing a $\pi^{0}$ are vetoed with the requirement of $E_{\rm max} <0.3$~GeV and $E_{\rm sum} <0.5$~GeV,
where $E_{\rm max}$ and $E_{\rm sum}$ are the maximum energy and the energy sum of the unused 
showers, respectively.
The $\gamma^{\prime}$ signal is selected using the square of the recoil mass, 
$M^2_{\rm rec (\RecoilMiss)}$, against the ST $\LamCB$ and $p$.

After imposing all selection conditions mentioned
above, the distribution of $M^2_{\rm rec (\RecoilMiss)}$ of the accepted DT candidate events 
from the combined seven data samples at different CM energies is shown in Fig.~\ref{fig:recmiss}(a).
There is a peaking structure at the $K_L^0$ mass position, from the process 
 $\LamC \to pK_L^0$.

\section{Background analysis}~\label{sec:bkg}
\begin{figure}[!htp]
    \begin{center}
            \includegraphics[width=0.4\textwidth]{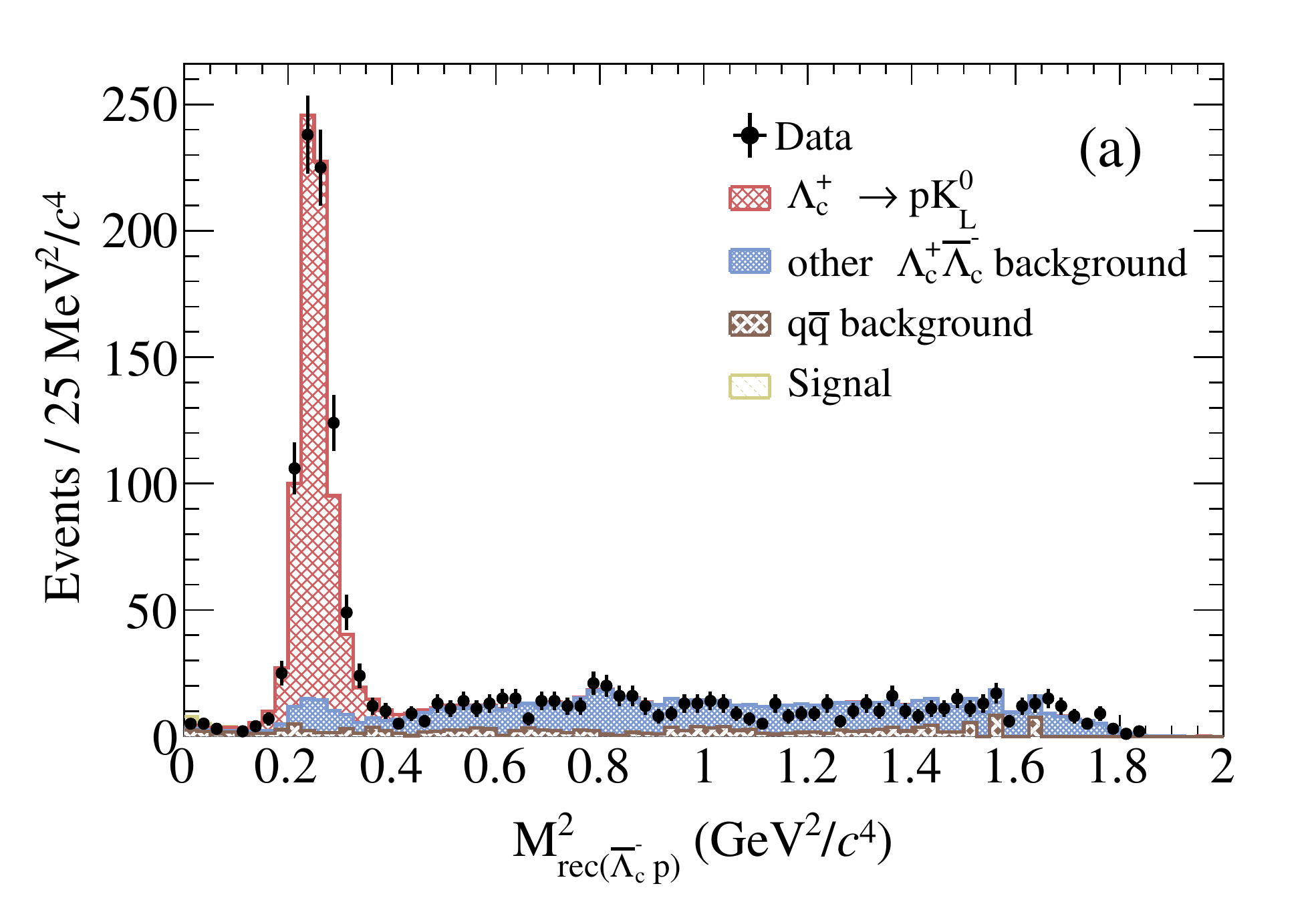}\label{fig:bkg_data}
            \includegraphics[width=0.4\textwidth]{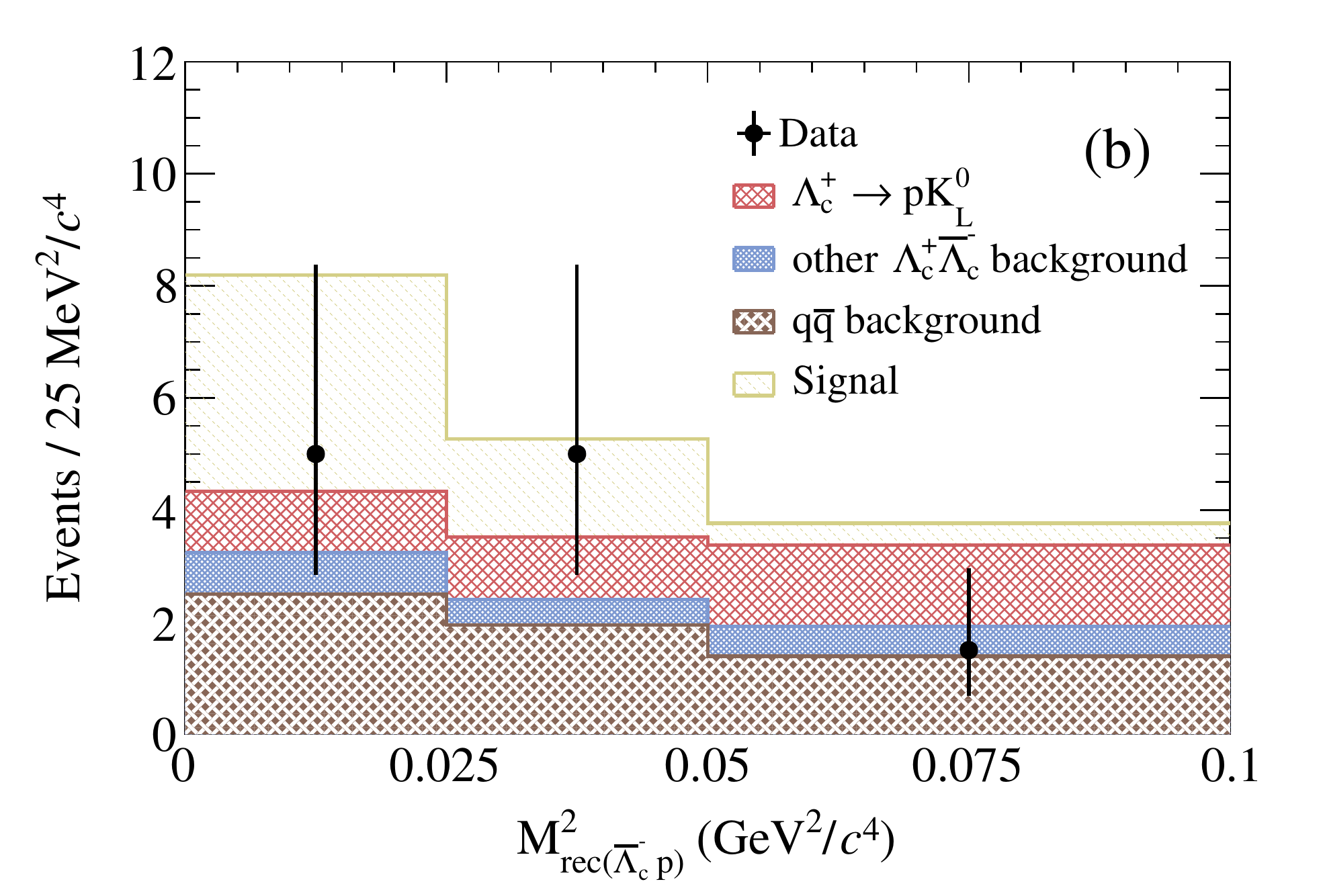}\label{fig:bkg_data_SR}
    \end{center}
        \caption{ (a) The full spectrum of $M^2_{\rm rec (\RecoilMiss)}$ of the accepted DT candidate events from the combined seven data samples.
              (b) The spectrum of $M^2_{\rm rec (\RecoilMiss)}$ of the accepted DT candidate events from the combined data in the signal region. 
              The black points with error bars are data; no events have been observed in the last bin.
              The red hatched histogram indicates $pK_{L}^{0}$ background.
              The blue hatched histogram is the $\LCpair$ background excluding $\LamC \to pK_{L}^{0}$ process.
              The brown hatched histogram represents the $q\bar{q}$ background.
              The yellow hatched histogram is the $\LamCPI$ signal, which is normalized to data luminosity with the upper limit of the BF.}
    \label{fig:recmiss}
\end{figure}

The potential background can be classified into two categories: those directly originated
from continuum hadron production in the $\ee$ annihilation, denoted as $q\bar{q}$ background, and those from the $\ee\to\LCpair$ events, denoted as $\LCpair$ background. The distribution and magnitude of $\LCpair$
background are estimated with the inclusive MC samples, where the peaking background $\LamC \to pK_L^0$ 
is extracted separately from the inclusive MC samples.
The $\LamC \to pK_L^0$ rate is normalized to the known BF~\cite{PDG:2020} 
and the remaining backgrounds are normalized to $\LCpair$ data yields.  
The $q\bar{q}$ background is investigated with $\mBC$ sideband region (2.21,2.26)~GeV/$c^2$ of ST
candidates in data, which is than reweighted to agree with those of the data in signal region.
The $q\bar{q}$ and $\LCpair$ backgrounds are estimated to be $7.4 \pm 0.4$ and $7.2 \pm 1.4$, respectively.
The resultant $M^2_{\rm rec (\RecoilMiss)}$ distribution is depicted in Fig.~\ref{fig:recmiss}(a).

\section{upper limit setting}
In order to extract the signal yield, a signal region is defined as (0.0, 0.1) $\gevcs$ in the $M^2_{\rm rec (\RecoilMiss)}$ distribution,
corresponding to a 96\% signal detection efficiency after imposing all the selection criteria. 
The distribution of $M^2_{\rm rec (\RecoilMiss)}$ in the signal region is shown in Fig.~\ref{fig:recmiss}(b).
Thirteen candidate events are observed in the signal region, while the background events are estimated to be $14.6 \pm 1.5$, where the uncertainty is statistical only.
The $\LamC$ decay BF is determined as in Eq.~(\ref{eq:br}).
The detection efficiency $\epsilon_{ij}^{\mathrm{ST}}$ is obtained with the same procedure 
as in Ref.~\cite{Ablikim:2022SCS}, and the $\epsilon_{ij}^{\mathrm{DT}}$ is derived with the signal MC samples.  
The DT efficiencies are summarized in \tablename~\ref{tab:efficiency-DT-signal}.

\begin{table}[!htbp]
  \begin{center}
  \caption{ 
           The DT detection efficiencies in percentage for ten tag modes and seven data samples at different CM energies.
            The statistical uncertainties are lower than 0.3\%.}
    \begin{tabular}{ l c c c c c c c}
      \hline
      \hline
         $\sqrt{s}$ (GeV)   &    $4.600$   &  $4.612$   &   $4.628$   &    $4.641$   &   $4.661$   &   $4.682$   &  $4.699$  \\
      \hline

            $\Bpkpi$                       &  34.3   &  33.8   &  32.3  &  32.0  &  31.6  & 31.3  & 30.7  \\
            $\Bpks$                        &  39.1   &  37.1   &  34.6  &  33.6  &  33.9  & 32.8  & 31.8  \\
            $\bar{\Lambda}\pi^-$           &  32.9   &  30.5   &  29.1  &  29.5  &  28.5  & 26.9  & 25.4  \\
            $\Bpkpi\pi^0$                  &  12.9   &  12.4   &  12.3  &  12.6  &  11.9  & 11.9  & 11.5  \\
            $\Bpks\pi^0$                   &  14.9   &  14.2   &  13.5  &  13.4  &  13.4  & 13.5  & 12.7  \\
            $\bar{\Lambda}\pi^-\pi^0$      &  13.6   &  12.7   &  12.2  &  12.3  &  12.0  & 11.6  & 11.2  \\
            $\Bpks\pi^+\pi^-$              &  15.4   &  14.5   &  13.7  &  13.6  &  13.4  & 13.3  & 13.2  \\
            $\bar{\Lambda}\pi^-\pi^+\pi^-$ &  10.4   &   9.8   &   9.3  &   9.6  &   9.3  &  9.3  &  9.1  \\
            $\bar{\Sigma}^0\pi^-$          &  18.8   &  17.7   &  17.1  &  16.2  &  14.7  & 14.6  & 15.0  \\
            $\bar{\Sigma}^-\pi^+\pi^-$     &  17.5   &  16.5   &  16.3  &  16.2  &  15.8  & 15.0  & 14.9  \\
      \hline\hline
    \end{tabular}
    \label{tab:efficiency-DT-signal}
  \end{center}
\end{table}

\begin{figure}[!htp]
    \begin{center}
        \includegraphics[width=0.4\textwidth]{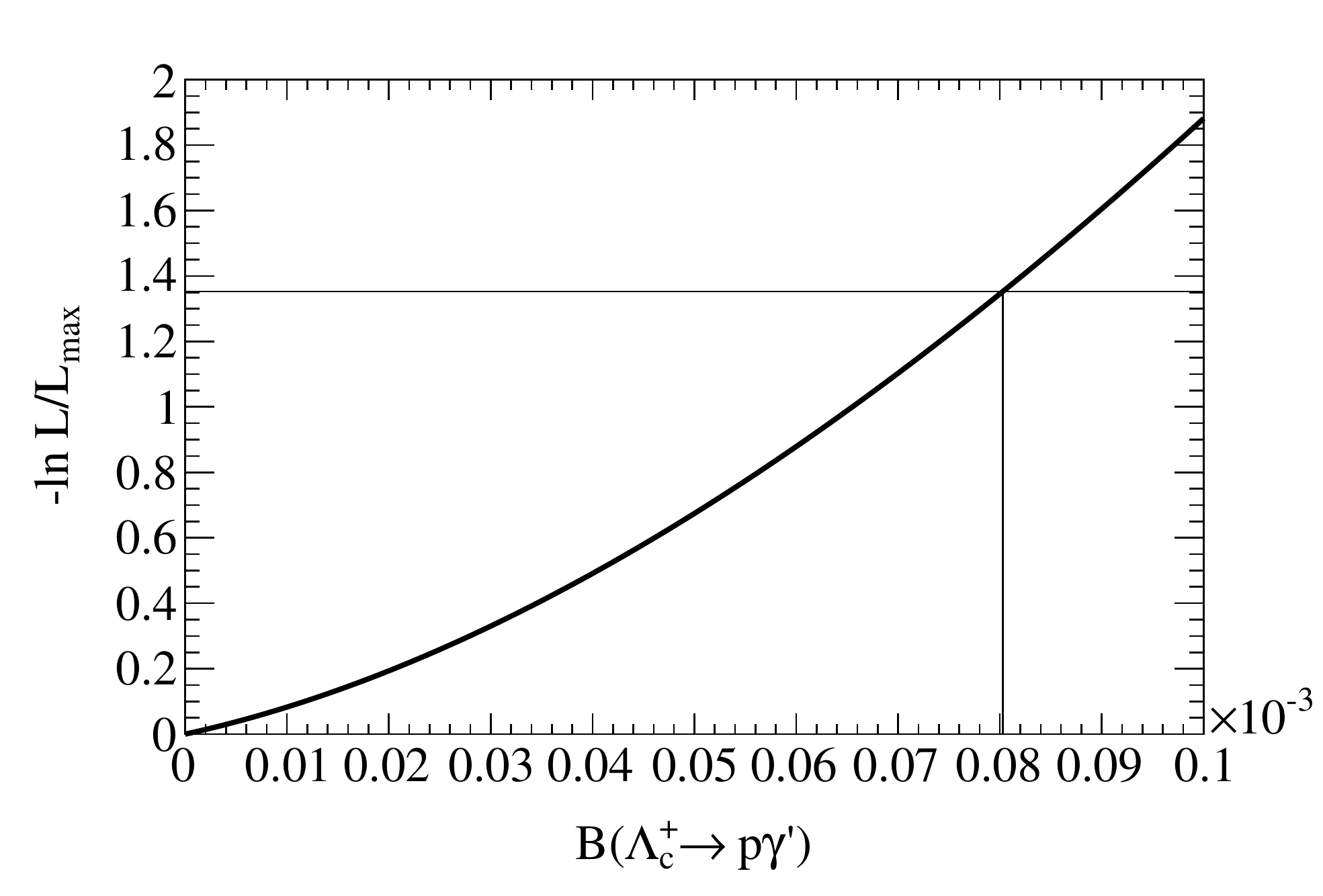}
    \end{center}
    \caption{
       The profile-likelihood curve versus $\mathcal{B}(\LamCPI)$. The black solid curve is the scan result with systematic uncertainties.
       The cross of the curve indicates the upper limit of the BF at 90\% CL. }
     \label{fig:limit_scan}
\end{figure}

Since no significant signal is observed, the profile-likelihood
approach~\cite{Cowan:LH} is used to determine the upper limit on the BF of $\LamCPI$.
The likelihood function which depends on the parameter of interest $\mathcal{B}(\LamCPI)$ and the nuisance
parameters $\theta = (\epsilon_{\rm eff}, N_{\rm bkg})$ is defined as:

\begin{equation} \label{eq:likelihood}
  \mathcal{L}(\mathcal{B}(\LamCPI), \theta) = P(N_{\rm obs}|N_{\rm exp}) \cdot G(\theta) 
\end{equation}
where the observed events are assumed to follow a Poisson
distribution ($P$). The $N_{\rm exp}$ is the expected number of events; 
it is defined as the sum of the number of background events and number 
of signal events estimated in the signal region,
corresponding to Eq.~(\ref{eq:br}).  
The detection efficiency $\epsilon_{\rm eff}$ and $N_{\rm bkg}$ follow Gaussian distributions ($G$).
The upper limit on the BF of $\LamCPI$ is determined by scanning the parameter of interest.
The resultant profile-likelihood scan distribution is presented in Fig.~\ref{fig:limit_scan}.
The upper limit is calculated to be $\mathcal{B}(\LamCPI) < 8.0\times 10^{-5}$ at 90\% confidence level (CL),
where the statistical and systematic uncertainties are all incorporated.
The systematic uncertainties associated with the detection efficiency and
the background estimation are performed with the two nuisance parameters in a profile-likelihood fit.

\section{Systematic Uncertainty}

The systematic uncertainties for the BF measurement include those associated with the ST yields ($\Nsingle$),
reconstruction efficiencies of the ST $\LamCB$ ($\effsingle$) and reconstruction efficiencies of the DT ($\effdouble$).
As the DT technique is adopted, the systematic uncertainties originating from reconstructing the ST side largely cancel.
Table~\ref{tab:sys-sum} summarizes the possible sources of systematic uncertainties.
Each of them is evaluated relative to the measured BF. 
The details are described below.

The uncertainties associated with the proton tracking and PID efficiencies are 
determined with the control sample $J/\psi \to p\bar{p} \pi^+\pi^-$~\cite{Uncer:pidTracking}.
The systematic uncertainties in proton tracking and PID are assigned to be 1.0\% each.

The uncertainty in the ST yields is 0.5\%, which arises from the statistical uncertainty and 
fitting $\mBC$ distributions. The uncertainty in the fitting procedure is evaluated by varying the 
ARGUS background parameter and changing the Gaussian function to a sum of two Gaussian functions. 

The systematic uncertainty for the requirements of $E_{\rm max}$ and $E_{\rm sum}$ is
studied using the control sample of $\Lambda_{c}^{+} \to pK^{-}\pi^{+}$. 
The systematic uncertainty, 3.0\%, is determined by comparing the efficiencies between data and MC simulation.

According to Eq.~(\ref{eq:br}), the uncertainty related to the ST
efficiency is expected to be canceled. However, due to the
different multiplicities of tracks and showers in the rest of the event, 
the ST efficiencies estimated with
the generic and the signal MC samples are expected to
differ slightly. Thus, the uncertainty associated with the ST
efficiency is not canceled fully, which results in a so called
“tag bias” uncertainty. 
The difference of ST efficiency between generic and the signal MC samples, 0.9\%,
is assigned as the corresponding uncertainty.

The uncertainty in the $q\bar{q}$ background estimation is obtained by widening the $\mBC$ sideband range 
by 5 MeV/$c^2$ in comparison to the nominal one, yielding 4.1\%.
In addition, the statistical uncertainty of 5.7\% on the data yield in the $\mBC$ sideband range (2.21,2.26)~GeV/$c^2$ is taken into account. 
The total $q\bar{q}$ background systematic uncertainty of 7.0\% is the quadratic sum of these two effects.  

The main $\LCpair$ background is the $\LamC \to pK_{L}^{0}$ process.
The uncertainty of its BF quoted from the PDG on ${\cal B}(\LamC \to pK_{S}^{0})$~\cite{PDG:2020} is 5.0\%, 
and the fraction of $pK_{L}^{0}$ in the signal region is 36\%.  The net 
systematic uncertainty in $\LCpair$ background estimation is thus 1.8\%.

Other uncertainties are negligible. 
Assuming that all the sources of the uncertainties are uncorrelated, 
the total systematic uncertainties associated with the detection efficiency and
the background estimation are 3.5\% and 7.2\%, respectively.

\begin{table}[htbp]
  \begin{center}
  \caption{The systematic uncertainties in percentage for $\LamCPI$.}\label{tab:sys-sum}
    \begin{tabular}{ l  c }
      \hline
      \hline
          mode   &    uncertainty (\%) \\
      \hline
            $p$ tracking                                       & 1.0 \\
            $p$ PID                                            & 1.0 \\
            ST yield                                   & 0.5 \\
            $E_{\rm max}$ and $E_{\rm sum}$ requirements       & 3.0 \\
            Tag bias                                           & 0.9 \\
       \hline
            $q\bar{q}$ background estimation                   & 7.0 \\
            $\LCpair$ background estimation                    & 1.8 \\ 
       \hline\hline
    \end{tabular}
  \end{center}
\end{table}

\section{Summary}
\label{sec:summary}

In summary, with a sample of 4.5~$\invfb$ collected at 
CM energies between 4.600 and 4.699~GeV with the BESIII detector,
the first investigation for a massless dark photon in $\LamCPI$ decay is carried out.
No significant signal is observed with respect to the expected background. The upper limit on 
the BF of $\LamCPI$ is measured to be $\mathcal{B}(\LamCPI) < 8.0\times 10^{-5}$ at 90\% CL. 
It is below the sensitivity of theory prediction in Ref.~\cite{Tandean:cFCNC}, which predicts the BF to be $1.6\times 10^{-5}$ or $9.1\times 10^{-6}$
with different inputs of form factors.
A more stringent constrain on $\mathcal{B}(\LamCPI)$ is expected in the near future with larger $\Lambda_c^+$ samples at BESIII~\cite{Ablikim:2019hff}.

\section*{ACKNOWLEDGEMENTS}

The BESIII collaboration thanks the staff of BEPCII, the IHEP computing center and the supercomputing center of the
University of Science and Technology of China (USTC) for their strong support.
Authors are grateful to Jusak Tandean for enlightening discussions.
This work is supported in part by National Key R\&D Program of China under Contracts Nos. 2020YFA0406400, 2020YFA0406300; 
National Natural Science Foundation of China (NSFC) under Contracts 
No. 11635010, No. 11735014, No. 11835012, No. 11935015, No. 11935016, 
No. 11935018, No. 11961141012, No. 12022510, No. 12025502, No. 12035009, No. 12035013, 
No. 12192260, No. 12192261, No. 12192262, No. 12192263, No. 12192264, No. 12192265, 
No. 12005311; 
the Fundamental Research Funds for the Central Universities, Sun Yat-sen University, University of Science and Technology of China;
100 Talents Program of Sun Yat-sen University;
the Chinese Academy of Sciences (CAS) Large-Scale Scientific Facility Program; 
Joint Large-Scale Scientific Facility Funds of the NSFC and CAS under Contract No. U1832207; 
100 Talents Program of CAS; 
The Institute of Nuclear and Particle Physics (INPAC) and Shanghai Key Laboratory for Particle Physics and Cosmology; 
ERC under Contract No. 758462; 
European Union's Horizon 2020 research and innovation programme under Marie Sklodowska-Curie grant agreement under Contract No. 894790; 
German Research Foundation DFG under Contracts Nos. 443159800, Collaborative Research Center CRC 1044, GRK 2149; 
Istituto Nazionale di Fisica Nucleare, Italy; 
Ministry of Development of Turkey under Contract No. DPT2006K-120470; 
National Science and Technology fund; 
National Science Research and Innovation Fund (NSRF) via the Program Management Unit for Human Resources \& Institutional Development, 
Research and Innovation under Contract No. B16F640076; 
STFC (United Kingdom); 
Suranaree University of Technology (SUT), Thailand Science Research and Innovation (TSRI), 
and National Science Research and Innovation Fund (NSRF) under Contract No. 160355; 
The Royal Society, UK under Contracts Nos. DH140054, DH160214; 
The Swedish Research Council; 
U. S. Department of Energy under Contract No. DE-FG02-05ER41374.


\begin{thebibliography}{99}

\bibitem{Glashow:fcnc}
S. L. Glashow, J. Iliopoulos, and L. Maiani, 
\href{https://doi.org/10.1103/PhysRevD.2.1285}{Phys. Rev. D {\bf 2} 1285, (1970).}

\bibitem{Tanabashi:charmFCNC}
M. Tanabashi {\it et al.} (Particle Data Group),
\href{https://doi.org/10.1103/PhysRevD.98.030001}{Phys. Rev. D {\bf 98}, 030001 (2018).}

\bibitem{Aulakh:model1}
C. S. Aulakh and R. N. Mohapatra, 
\href{https://doi.org/10.1016/0370-2693(82)90262-3}{Phys. Lett. B {\bf 119}, 136 (1982).}

\bibitem{Glashow:model2}
S. Glashow and S. Weinberg, 
\href{https://doi.org/10.1103/PhysRevD.15.1958}{Phys. Rev. D {\bf 15}, 1958 (1977).}

\bibitem{dark:mode1}
B. Holdom, 
\href{https://doi.org/10.1016/0370-2693(86)91377-8}{Phys. Lett. B {\bf 166}, 196 (1986).}

\bibitem{Dobrescu:dark1}
B. A. Dobrescu, 
\href{https://doi.org/10.1103/PhysRevLett.94.151802}{Phys. Rev. Lett. {\bf 94}, 151802 (2005).}

\bibitem{Dobrescu:dark2}
E. Gabrielli, B. Mele, M. Raidal, and E. Venturini, 
\href{https://doi.org/10.1103/PhysRevD.94.115013}{Phys. Rev. D {\bf 94}, 115013 (2016).}

\bibitem{dark:mode2}
M. Fabbrichesi, E. Gabrielli, and B. Mele, 
\href{https://doi.org/10.1103/PhysRevLett.119.031801}{Phys. Rev. Lett. {\bf 119}, 031801 (2017).}

\bibitem{Ying:darkphoton}
J. Y. Su and J. Tandean, 
\href{https://doi.org/10.1103/PhysRevD.101.035044}{Phys. Rev. D {\bf 101}, 035044 (2020).}

\bibitem{Ackerman:intre1}
L. Ackerman, M. R. Buckley, S. M. Carroll, and M. Kamionkowski, 
\href{https://doi.org/10.1103/PhysRevD.79.023519}{Phys. Rev. D {\bf 79}, 023519 (2009).}

\bibitem{Pan:intre2}
J. X. Pan, M. He, X. G. He, and G. Li, 
\href{https://doi.org/10.1016/j.nuclphysb.2020.114968}{Nucl. Phys. B {\bf 953}, 114968 (2020).}

\bibitem{Barger:intre3}
V. Barger, C. W. Chiang, W. Y. Keung, and D. Marfatia, 
\href{https://doi.org/10.1103/PhysRevLett.108.081802}{Phys. Rev. Lett. {\bf 108}, 081802 (2012).} 

\bibitem{Chiang:intre4}
C. W. Chiang and P. Y. Tseng, 
\href{https://doi.org/10.1016/j.physletb.2017.02.022}{Phys. Lett. B {\bf 767}, 289 (2017).}

\bibitem{Batley:intre5}
J. R. Batley {\it et al.} (NA48/2 Collaboration),
\href{https://doi.org/10.1016/j.physletb.2015.04.068}{Phys. Lett. B {\bf 746}, 178 (2015).} 

\bibitem{Pospelov:intre6}
M. Pospelov, 
\href{https://doi.org/10.1103/PhysRevD.80.095002}{Phys. Rev. D {\bf 80}, 095002 (2009).}

\bibitem{LHCb:lepvio}
R.~Aaij {\em et~al.} (LHCb Collaboration),
\href{https://doi.org/10.1038/s41567-021-01478-8}{Nat. Phys. {\bf 18}, 277 (2022).}

\bibitem{BESIII:etainv}
M.~Ablikim {\em et~al.} (BESIII Collaboration), 
\href{https://doi.org/10.1103/PhysRevD.87.012009}{Phys. Rev. D {\bf 87},  012009 (2013).}

\bibitem{BESIII:omegainv}
M.~Ablikim {\em et~al.} (BESIII Collaboration), 
\href{https://doi.org/10.1103/PhysRevD.98.032001}{Phys. Rev. D {\bf 101}, 032001 (2018).}

\bibitem{BESIII:Laminv}
M.~Ablikim {\em et~al.} (BESIII Collaboration), 
\href{https://doi.org/10.1103/PhysRevD.105.L071101}{Phys. Rev. D {\bf 105}, L071101 (2022).}

\bibitem{BESIII:Jpsiinv}
M.~Ablikim {\em et~al.} (BESIII Collaboration), 
\href{https://doi.org/10.1103/PhysRevD.101.112005}{Phys. Rev. D {\bf 101}, 112005 (2020).}

\bibitem{Tandean:cFCNC}
J. Y. Su and J. Tandean, 
\href{https://doi.org/10.1103/PhysRevD.102.115029}{Phys. Rev. D {\bf 102}, 115029 (2020).}

\bibitem{MarkIII:DT}
J. Adler {\it et al.},
\href{https://doi.org/10.1103/PhysRevLett.62.1821}{Phys. Rev. Lett. {\bf 62}, 1821 (1989).}

\bibitem{Ablikim:2009aa} 
M.~Ablikim {\em et~al.} (BESIII Collaboration), 
\href{https://doi.org/10.1016/j.nima.2009.12.050}{Nucl. Instrum. Meth. A {\bf 614}, 345 (2010).}

\bibitem{Ablikim:2019hff}
M.~Ablikim {\em et~al.} (BESIII Collaboration), 
\href{https://doi.org/10.1088/1674-1137/44/4/040001}{Chin. Phys. C {\bf 44}, 040001 (2020).}

\bibitem{LiGuo:MRPC}
X. Li {\em et al.},  \href{https://doi.org/10.1007/s41605-017-0014-2}{Radiat. Detect. Technol. Methods {\bf 1}, 13 (2017);}
Y. X. Guo {\em et al.}, \href{https://doi.org/10.1007/s41605-017-0012-4}{Radiat. Detect. Technol. Methods {\bf 1}, 15 (2017).}

\bibitem{BESIII:lumi0}
M. Ablikim {\em et al.} (BESIII Collaboration), 
\href{https://doi.org/10.1088/1674-1137/40/6/063001}{Chin. Phys. C {\bf 40}, 063001 (2016).}

\bibitem{BESIII:lumi1}
M. Ablikim {\em et al.} (BESIII Collaboration), arXiv:2203.03133.

\bibitem{BESIII:lumi2}
M. Ablikim {\em et al.} (BESIII Collaboration), arXiv:2205.04809.

\bibitem{Agostinelli:2002hh}
S. Agostinelli {\it et al.},
\href{https://doi.org/10.1016/S0168-9002(03)01368-8}{Nucl. Instrum. Meth. A {\bf 506}, 250 (2003).}

\bibitem{Jadach:2000ir}
S. Jadach {\it et al.},
\href{https://doi.org/10.1103/PhysRevD.63.113009}{Phys. Rev. D {\bf 63}, 113009 (2001).}

\bibitem{Lange:2001uf}
D. J. Lange, 
\href{https://doi.org/10.1016/S0168-9002(01)00089-4}{Nucl. Instrum. Meth. A {\bf 462}, 152 (2001).}

\bibitem{Ping:2008zz}
R. G. Ping, 
\href{https://doi.org/10.1088/1674-1137/32/8/001}{Chin. Phys. C {\bf 32}, 599 (2008).}

\bibitem{PDG:2020}
P. A. Zyla {\it et al.} (Particle Data Group), 
\href{https://doi.org/10.1093/ptep/ptaa104}{Prog. Theor. Exp. Phys. {\bf 2020}, 083C01 (2020) and 2021 update.}

\bibitem{Chen:2000tv}
J. C. Chen, G. S. Huang, X. R. Qi, D. H. Zhang, and Y. S. Zhu, 
\href{https://doi.org/10.1103/PhysRevD.62.034003}{Phys. Rev. D {\bf 62}, 034003 (2000).}

\bibitem{YANG:2014}
R. L. Yang, R. G. Ping, H Chen, 
\href{https://doi.org/10.1088/0256-307X/31/6/061301}{Chin. Phys. Lett. {\bf 31}, 061301 (2014).}

\bibitem{Richter-Was:1992hxq}
R. W. Elzbieta, 
\href{https://doi.org/10.1016/0370-2693(93)90062-M}{Phys. Lett. B {\bf 303}, 163 (1993).}

\bibitem{ARGUS:1990hfq}
H. Albrecht {\it et al.},
\href{https://doi.org/10.1016/0370-2693(90)91293-K}{Phys. Lett. B {\bf 241}, 278 (1990).}

\bibitem{Ablikim:2022SCS}
M.~Ablikim {\em et~al.} (BESIII Collaboration), 
\href{https://doi.org/10.1103/PhysRevLett.128.142001}{Phys. Rev. Lett. {\bf 128}, 142001 (2022).}

\bibitem{Cowan:LH}
G. Cowan, K. Cranmer, E. Gross and O. Vitells, 
\href{https://doi.org/10.1140/epjc/s10052-011-1554-0}{Eur. Phys. J. C {\bf 71}, 1554 (2011).} 

\bibitem{Uncer:pidTracking}
M. Ablikim {\it et al.} (BESIII Collaboration), 
\href{https://doi.org/10.1088/1674-1137/40/2/026201}{Chin. Phys. C {\bf 40}, 026201 (2016).}





\end{thebibliography}
\end{document}